\documentclass[fleqn,twoside]{article}
\usepackage{espcrc2}

\usepackage{graphicx}
\usepackage[figuresright]{rotating}

\newcommand{\AmS}{{\protect\the\textfont2
  A\kern-.1667em\lower.5ex\hbox{M}\kern-.125emS}}

\title{
 Higher orders of the high-temperature expansion for the Ising model
 in three dimensions
}

\author{H. Arisue\address[OPCT]{Osaka Prefectural College of Technology, 
   Saiwai-cho 26-12, Neyagawa, Osaka 572-8572, Japan}
   \thanks{email address:arisue@las.osaka-pct.ac.jp},
   T. Fujiwara\address[KU]{Faculty of General Studies, Kitasato University, 
   Kitasato 1-15-1, Sagamihara, Kanagawa 228-8555, Japan}
   \thanks{email address:fujiwara@clas.kitasato-u.ac.jp}
     and
   K. Tabata\addressmark[OPCT]\thanks{email address:tabata@las.osaka-pct.ac.jp}
        }
       
\begin{document}
\begin{abstract}
 The new algorithm of the finite lattice method is applied to generate 
the high-temperature expansion series of the simple cubic Ising model 
to $\beta^{50}$ for the free energy, 
to $\beta^{32}$ for the magnetic susceptibility and
to $\beta^{29}$ for the second moment correlation length.
The series are analyzed to give the precise value of the critical point
and the critical exponents of the model.
\vspace{1pc}
\end{abstract}

\maketitle

%%%%%%%%%%%%%%%%%%%%%%%%%%%%%%%%%%%%%%%%%%%%%%%%%%%%%%%%%%%%%%%%%%%%%%%
\section{INTRODUCTION}
The finite lattice method\cite{Enting1977,Arisue1984,Creutz1991} is a powerful 
tool to generate the exact high- and low-temperature series and other exact 
expansion series for the spin models in the infinite volume limit.
Using this method we can skip the job of 
listing up all the relevant diagrams and of counting the number they appear,
which is inevitable in the graphical method. 
Then the main task is reduced to the calculation of the partition 
function for the appropriate finite size lattices,
which can be done efficiently by the site-by-site 
integration using the transfer matrix formalism\cite{Enting1980,Bhanot1990} 
without the graphical technique.
The CPU time and the computer memory 
needed to obtain the series to order $N$ increase exponentially with $N$
in two dimensions, 
while they grow up exponentially with $N^2$ in three dimensions.
Thus the finite lattice method has been extremely effective in two dimensions,
but it has generated the series only in some limited cases in three dimensions.

Recently we presented the new algorithm 
of the finite lattice method\cite{Arisue2003a,Arisue2003b} 
in which the CPU time and the computer memory increase exponentially 
with $N\log{N}$.
It enabled us to extend the high-temperature expansion series 
for the free energy of the simple cubic Ising model to order $\beta^{46}$ 
from the previous series of order $\beta^{26}$, which was given by the previous
algorithm of the finite lattice method, and to order $\beta^{32}$ 
for the magnetic susceptibility from the previous series of order $\beta^{25}$, 
which was given by the graphical method.

Here we apply the method newly to the high-temperature expansion of 
the second moment correlation length to order $\beta^{29}$ and 
also extend the series for the free energy to order $\beta^{50}$.
The series themselves will be presented elsewhere\cite{Arisue2003c} 
and we give here the result of the analysis of the newly generated
series combined with the series for the 
magnetic susceptibility already obtained.

%%%%%%%%%%%%%%%%%%%%%%%%%%%%%%%%%%%%%%%%%%%%%%%%%%%%%%%%%%%%%%%%%%%%%%%
\section{SUSCEPTIBILITY AND CORRELATION LENGTH}
The magnetic susceptibility $\chi$ and 
the second moment correlation length $\xi_2$ are expected to behave like
\begin{eqnarray} 
\chi &=& A_1\; (\beta_c-\beta)^{-\gamma} 
     \left[ 1 + c_1(\beta_c-\beta)^{\Delta}
      + \cdots \ \ \right] 
    \nonumber
\end{eqnarray} 
\begin{eqnarray} 
{\xi_2}^2 &=& A_2\; (\beta_c-\beta)^{-2\nu} 
     \left[ 1 + c_2(\beta_c-\beta)^{\Delta}
      + \cdots \ \ \right] 
    \nonumber
\end{eqnarray} 
Here $\beta_c$ is the critical point and $\Delta$ is 
the leading correction-to-scaling exponent.
%, which is expected to be $\Delta \sim 0.5$.

\begin{figure}[tb]
\includegraphics{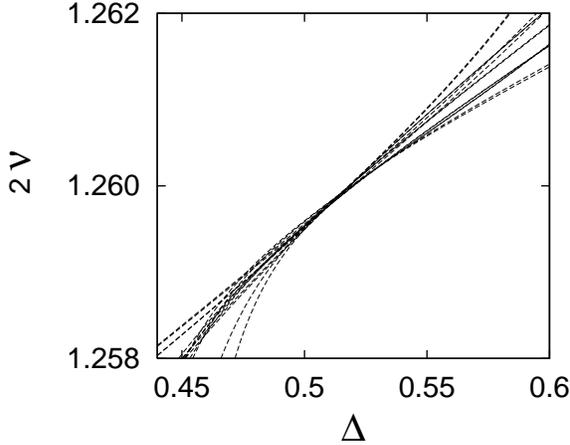}
\vspace{-9mm}
\caption{Critical exponent $2\nu$ plotted versus $\Delta$ in the M2 method of the series
analysis for ${\xi_2}^2$. 
}
\vspace{-3mm}
\label{fig:m2}
\end{figure}

There are several evaluation of the value of $\Delta$ to be around $0.5$.
To make more precise evaluation we plot $2\nu$ versus $\Delta$ in Figure \ref{fig:m2}
in the analysis of the $\xi_2$ series by the M2 method\cite{Adler1983}.
The narrow crossing region of the lines gives $\Delta=0.51(1)$.

The modified ratio series\cite{ZinnJustin1981}
\begin{eqnarray}
 (\beta_c)_n \equiv \left(\frac{a_{n-2}a_{n-3}}{a_{n}a_{n-1}}\right) 
         \exp{\left[\frac{(s_{n} + s_{n-2})}{2s_{n}(s_{n} -s_{n-2})}\right] }
                 \label{eqn:defbetacn}    %\nonumber
\end{eqnarray} 
with
\begin{eqnarray}
 s_n = \left( \ln{(\frac{a_{n-2}^2}{a_{n}a_{n-4}}})^{-1}
            + \ln{(\frac{a_{n-3}^2}{a_{n-1}a_{n-5}}})^{-1}  \right)/2 \;
                     \nonumber
\end{eqnarray} 
made of the expantion coefficients $a_n$ for each quantity
 is expected to behave like 
\begin{eqnarray}
      (\beta_c)_n =
           \beta_c + \frac{b}{n^{1+\Delta}} 
              + \cdots \;. 
                     \label{eqn:betacn}
\end{eqnarray} 
Although the modefied ratio series for the longest series of each 
of the $\chi$ and ${\xi_2}^2$ 
can be fitted well by the first two terms in Eq. (\ref{eqn:betacn})
with $\Delta=0.51$, there is still a considerably large ambiguity 
in extrapolating the plot to $n\rightarrow \infty$.

\begin{figure}[tb]
\includegraphics{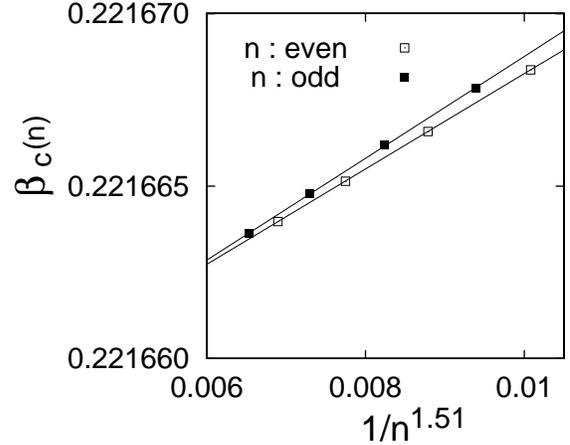}
\vspace{-9mm}
\caption{Modified ratio series for the combination of $\chi$ and $\xi_2$ 
with $\omega=0.055$ plotted versus $1/n^{1+\Delta}$}
\vspace{-6mm}
\label{fig:betac-n}
\end{figure}

Then we take the following combined quantity of the susceptibility and the 
correlation length.
\begin{eqnarray} 
\chi/(\xi^2)^{\omega}&& \nonumber\\
&&\!\!\!\!\!\!\!\!\!\!\!\!\!\!\!\!\!\!\!\!\!\!\!\!\!\!\!\!\!\!\!\!\!
= A_3\; (\beta_c-\beta)^{-\gamma+2\omega\nu} 
     \left[ 1 + c_3(\beta_c-\beta)^{\Delta}
      + \cdots\ \right].
                     \label{eqn:combinationa}
%   \nonumber
\end{eqnarray} 
We can expect that, if we take an appropriate value of $\omega$, 
the third and later terms are very small and 
the ratio series (\ref{eqn:defbetacn}) for this quantity can be fitted 
quite well by the first two terms in Eq. (\ref{eqn:betacn}).
In fact we find as in Figure \ref{fig:betac-n} that the situation is realized 
for $\omega=0.055$ for even series of $n=22-28$ 
giving $\beta_c=0.2216545(1)$ if we assume $\Delta=0.51(1)$. 
Similarly the combined quantity ${\xi_2}^2/\chi^{\omega'}$ 
with $\omega'= -0.15$ gives $\beta_c=0.2216542(4)$ 
with the same value of $\Delta$ assumed.

There are two estimates for the critical point of the model 
by the recent large scale numerical simulations. One is $\beta_c=0.22165459(10)$
by Bl\"{o}te {\it et al.}\cite{Blote1999} 
using the cluster algorithm and the other is
$\beta_c=0.2216545(15)$\cite{Ito2002} by Ito {\it et al.} 
using the non-equilibrium relaxation method. They are contradictory 
to each other. Our estimation by the analysis of the high-temperature series
is consistent with these estimates.

As for the critical exponents the modified ratio series for 
the combination (\ref{eqn:combinationa}) of the magnetic susceptibility 
and the second moment correlation length 
gives $\gamma=1.2369(3)$ and $\nu=0.6298(3)$ for $\Delta=0.51(1)$.
These values are consistent with the estimate of $\gamma=1.2373(2)$
and $\nu=0.6301(2)$ by the high-temperature expansion of the generalized 
Ising model \cite{Campostrini2002} and of $\gamma=1.2396(13)$
and $\nu=0.6304(13)$ by the perturbative renormalization method\cite{Guida1997}.

\section{SPECIFIC HEAT}
For the critical exponent of the specific heat 
\begin{eqnarray} 
C_H &\equiv& \beta^2 \frac{d^2 f}{d\beta^2} \nonumber\\
  &= & A_4\; (\beta_c-\beta)^{-\alpha} 
     \left[ 1 + c_4 (\beta_c-\beta)^{\Delta}
      + \cdots \ \ \right] ,
    \nonumber
\end{eqnarray} 
the ratio series 
\begin{eqnarray}
 \alpha_n &\equiv& [\beta_c^2 \frac{a_{2n}}{a_{2n-2}} -1] n +1 
                     \nonumber\\
          &=  & \alpha + \frac{b}{n^{\Delta}} + \frac{c}{n} 
             + \cdots \ \ 
                     \nonumber
\end{eqnarray} 
with $n\le 50$ 
gives $\alpha=0.1035\pm 0.0005$ for $\beta_c=0.2216546(1)$ and $\Delta=0.51(1)$. 
This estimate of $\alpha$ is consistent with that 
by the inhomogeneous differential approximation for the high-temperature series.
From the hyperscaling relation $\alpha=2-d\nu$, 
the range of $\nu=0.6295-0.6305$, within which all of the recent precise 
estimates of the critical exponent for the correlation length falls,
gives $\alpha=0.1085-0.1115$.
Thus there is a discrepancy between the estimate of $\alpha$ 
from the high-temperature
series of the specific heat and the estimate from the hyperscaling relation.

%%%%%%%%%%%%%%%%%%%%%%%%%%%%%%%%%%%%%%%%%%%%%%%%%%%%%%%%%%%%%%%%%%%%%%%
\section{SUMMARY AND DISCUSSION}
We have calculated the high-temperature series for the second moment 
correlation length of the simple cubic Ising model 
using the new algorithm of the finite lattice method. 
Combining this series with the previously obtained series for the 
magnetic susceptibility, we have given precise estimates of the 
critical point and the critical exponents, which is consistent with other
recent estimates.
We have also extended the series for the free energy, which gives 
the estimate of the critical exponent for the specific heat, which is 
inconsistent with that derived from the critical exponent of the 
correlation length using the hyperscaling relation.

The new algorithm can also be applied to the high-temperature expansion 
of the models that have continuous spin variables 
such as the XY model in three dimensions. 
Especially it is important to calculate the high order series 
of the high-temperature expansion for the XY model, 
since the phase transition of this model 
belongs to the same universality class as the lambda transition 
of the helium and the specific heat exponent at this phase transition point
was observed very precisely in the Space Shuttle experiment\cite{Lipa1996}.

Furthermore the basic idea of the new algorithm 
can be used in the low-temperature expansion 
for the spin models.
We can expect that the new algorithm will enable us 
to generate the series for these models that are much longer 
than the presently available series.
%%%%%%%%%%%%%%%%%%%%%%%%%%%%%%%%%%%%%%%%%%%%%%%%%%%%%%%%%%%%%%%%%%%%%%%

\end{document}